\documentclass[aps,prl,twocolumn,superscriptaddress]{revtex4-2}

\usepackage{graphicx}
\usepackage{xcolor}
\usepackage{hyperref}

\begin{document}

\title{Transport spin polarization in RuO$_2$ films}

\author{Alexandra J. Howzen}
\affiliation{Department of Physics, Texas State University, San Marcos, Texas 78666, USA}

\author{Sachin~Gupta}
\affiliation{School of Physics and Astronomy, University of Leeds, Leeds, LS2 9JT, United Kingdom}

\author{Gavin Burnell}
\affiliation{School of Physics and Astronomy, University of Leeds, Leeds, LS2 9JT, United Kingdom}

\author{Nathan~Satchell}
\email{satchell@txstate.edu}
\affiliation{Department of Physics, Texas State University, San Marcos, Texas 78666, USA}

\begin{abstract}
Altermagnets host spin-split electronic bands without net magnetization, enabling spin-polarized transport in the absence of conventional ferromagnetism. RuO$_2$ has been proposed as a candidate altermagnet, yet experimental reports remain conflicting, particularly between bulk-sensitive probes and thin-film measurements. Here we investigate the electronic transport properties of epitaxial RuO$_2$ thin films using anomalous Hall effect measurements and point-contact Andreev reflection spectroscopy. We observe transport spin polarization and a strongly orientation-dependent anomalous Hall response, while magnetometry reveals no detectable net magnetization. The anomalous Hall effect appears only in ultrathin (110)-oriented films, consistent with symmetry-driven N\'eel-vector physics, and the measured transport spin polarization is systematically higher for (110)-oriented films than for (001)-oriented films, consistent with the crystallographic anisotropy of the spin-split bands. These results are consistent with altermagnetic behavior in RuO$_2$, with the experimentally accessible signatures confined to near-surface regions. They also establish superconducting transport spectroscopy as a metrology for identifying and characterizing altermagnet candidates.
\end{abstract}

\maketitle


Altermagnets are an emerging class of magnetic materials that merge key characteristics of ferromagnets and antiferromagnets: they host spin-polarized conduction bands and exhibit an anomalous Hall effect despite possessing antiparallel magnetic sublattices and zero net magnetization~\cite{smejkal_2022}. This behavior originates from symmetry-enforced spin splitting of the electronic structure, which enables ferromagnet-like transport responses in compensated magnetic systems~\cite{Bai_2024}. 

Ruthenium dioxide (RuO$_2$) has been proposed as a prototypical altermagnet candidate~\cite{vsmejkal2020crystal}. In its rutile structure, anisotropic Ru–O bonding breaks local inversion symmetry and is predicted to produce symmetry-enforced spin splitting of the Ru $4d$ bands. Yet, experiments have not reached consensus on the magnetic ground state~\cite{li2025exploration, hussain2025exploring}. Some experimental works, particularly those studying thin-films, report anomalous Hall effect signals~\cite{feng_2022, Jeong2025} and spin-split electronic states~\cite{fedchenko2024observation, lin2024observation}, while other works, particularly those studying bulk material, find no evidence of static magnetic order~\cite{kessler2024absence}. 
These observations suggest that magnetic order in RuO$_2$, if present, may be confined to near-surface regions or stabilized by symmetry reduction in thin films and strained heterostructures~\cite{akashdeep2025surfacelocalizedmagneticorderruo2, Jeong2025}. 

A direct measurement of the transport spin polarization in RuO$_2$ thin films would provide a quantitative benchmark for spin-polarized transport in candidate altermagnets. Establishing reliable values of spin polarization is essential for evaluating the potential of these systems in spintronic and superconducting spintronic devices, where many proposed architectures exploit the symmetry-driven spin-split band structures of altermagnets to realize magnetic spin valves, Josephson junctions, and related functionalities without net magnetization~\cite{PhysRevLett.126.127701,PhysRevB.108.174439,ouassou_2023,zhang_2023,beenakker_2023}.

In this Letter, we use point-contact Andreev reflection spectroscopy (PCAR)~\cite{soulen1998}, originally developed to measure spin polarization in ferromagnets~\cite{Strijkers_2001,Yates2018}, to probe spin-polarized transport in epitaxial RuO$_2$ thin films, yielding quantitative measurements of the transport spin polarization. Recent theoretical work has shown that Andreev reflection at altermagnet/superconductor interfaces can exhibit orientation-dependent, spin-polarized transport even in the absence of net magnetization, establishing PCAR as a suitable probe of symmetry-driven spin polarization in compensated magnetic systems~\cite{PhysRevB.108.054511,PhysRevB.108.L060508}. PCAR spectra from 60~nm RuO$_2$ films reveal transport spin polarization values that are systematically higher for (110)-oriented films than for (001)-oriented films at 4.2~K. In addition, we examine anomalous Hall transport in ultrathin 1.8~nm films and find a complementary orientation dependence: (110)-oriented films exhibit an anomalous Hall effect, while (001)-oriented films show no detectable anomalous Hall response, consistent with prior experimental reports and symmetry-based theoretical expectations for RuO$_2$~\cite{Jeong2025, vsmejkal2020crystal}.


Thin films of RuO$_2$ of thickness 60~nm and 1.8~nm were grown with the following orientations and corresponding substrates. (001) RuO$_2$ was deposited on (001) TiO$_2$/m-plane Al$_2$O$_3$ and m-plane Al$_2$O$_3$, and (110) RuO$_2$ was deposited on (110) TiO$_2$ and (001) MgO. For convenience, we define a sample identifier used throughout by the (orientation of RuO$_2$)/substrate or seed layer, e.g. (110)/TiO$_2$ corresponding to the (110) RuO$_2$ deposited on (110) TiO$_2$ substrate. 

Deposition was performed by reactive RF sputtering in a system with a base pressure of $1\times10^{-7}$~Torr from a 99.95\% pure Ru target in 99.9999\% pure Ar and O$_2$ process gasses at an elevated substrate temperature of 400$^\circ$C. The Ar and O$_2$ flow rates were 8 and 2.5~SCCM, respectively, for a growth pressure of approximately 1×10$^{-3}$~Torr.  An oxygen flow rate of 2.5~SCCM was chosen to maximize the amount of O$_2$ exposed to the growing film, since rutile RuO$_2$ is an oxygen rich phase, while avoiding the growth rate fluctuations from plasma instability observed when testing higher oxygen flow rates. The RF power was 72~W, giving a RuO$_2$ growth rate of 0.5~\AA/s.  To protect the RuO$_2$ between growth and PCAR measurements, a 3~nm Au capping layer was deposited after the sputter stage cooled to 25$^\circ$C. The Au layer was later removed via Ar ion milling prior to PCAR measurements. Additional details on the substrate preparation and TiO$_2$ buffer layer growth is available in the Supplemental Materials~\cite{SM}. 

Electrical transport measurements were performed in a variable temperature cryostat with a 9~T superconducting solenoid. Continuous film samples were contacted in the van der Pauw geometry and connected through a Keithley 7001 switch system to a Keithley 6221 current source and 2182A nanovoltmeter for resistivity and Hall effect measurements. An applied current of 100~$\mu$A was used for van der Pauw measurements. PCAR measurements followed established procedures~\cite{baltz2009conductance, PhysRevB.85.064410, Seemann_2010, 6971749, PhysRevB.76.174435, satchell2023}. Nb tips were fabricated from 99.9\% pure, 0.5~mm diameter Nb wire. Differential conductance was measured using an AC lock-in technique with Stanford Research Systems SR830 instruments. The tip position was controlled by a spring-loaded rod actuated with a micrometer screw. All PCAR measurements were carried out in liquid helium at 4.2~K and zero applied magnetic field. X-ray characterization was performed in a Rigaku SmartLab Diffractometer with a monochromated Cu k-$\alpha$ (1.5406~\AA) wavelength. Magnetization characterization was performed in a Quantum Design Magnetic Properties Measurement System (MPMS-3). 



\begin{figure}
    \centering
    \includegraphics[width=\columnwidth]{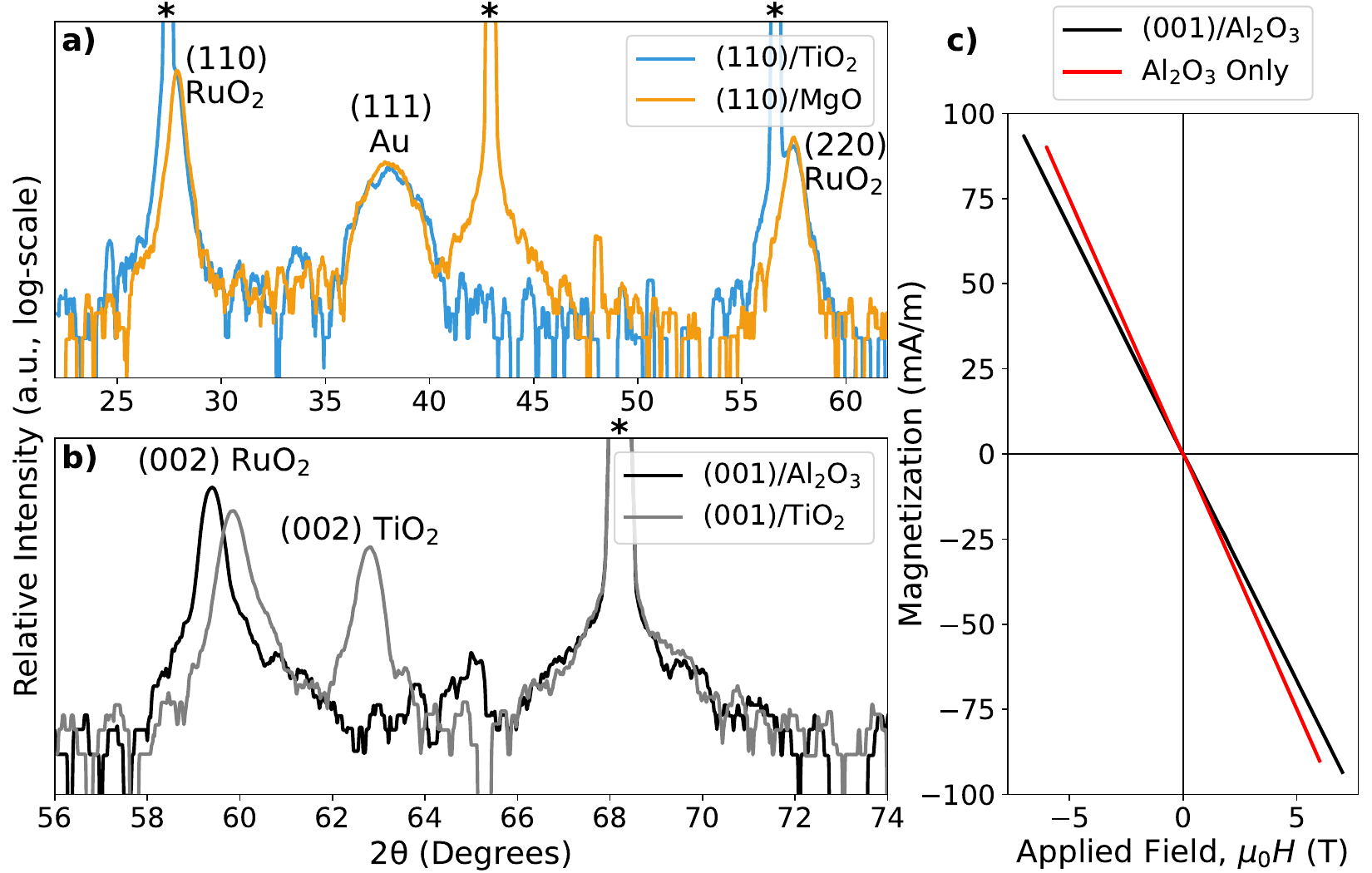}
    \caption{\label{Figure 1: XRD}
    Structural and magnetic properties of 60~nm RuO$_2$ thin films with a 3 nm Au cap. (a,b) X-ray diffraction of films deposited on four different substrates, (a) (110) oriented RuO$_2$ and (b) (001) oriented RuO$_2$ with their respective substrate peaks labeled (\textbf{$\ast$}). (c) SQUID magnetometry of the film deposited on Al$_2$O$_3$ and a bare substrate comparison with magnetization calculated from the measured areas and the nominal thickness of the film plus substrate. Magnetometry data acquired at 5~K for a magnetic field applied in-plane on a quartz paddle.}
\end{figure}

The structure of bulk, rutile RuO$_2$ (tetragonal P4$_2$/mnm) has crystal parameters $a = b = 4.492$~\AA, $c = 3.106$~\AA, and all angles are 90$^\circ$ \cite{10.1088/1361-648X/ade417}. We use X-ray diffraction to determine the structure of our 60~nm thick epitaxial thin films, shown in Fig.~\ref{Figure 1: XRD} (a,b). For growth on (110) TiO$_2$ and (100) MgO substrates, RuO$_2$ films have a preferred (110) orientation, and both (110) and (220) reflections are visible in Fig.~\ref{Figure 1: XRD} (a). The expected d-spacing corresponding to the (220) reflection is 1.589~\AA, which compares to the experimental values of 1.604~\AA~and 1.602~\AA~for the (110)/TiO$_2$ and (110)/MgO samples, respectively. These values correspond to small tensile strain (0.93\% and 0.81\%), indicating similar out-of-plane relaxation in the two (110)-oriented films. Additionally, the (111) peak of the 3~nm Au capping layer is observed.

\begin{figure*}
    \centering
    \includegraphics[width=1\linewidth]{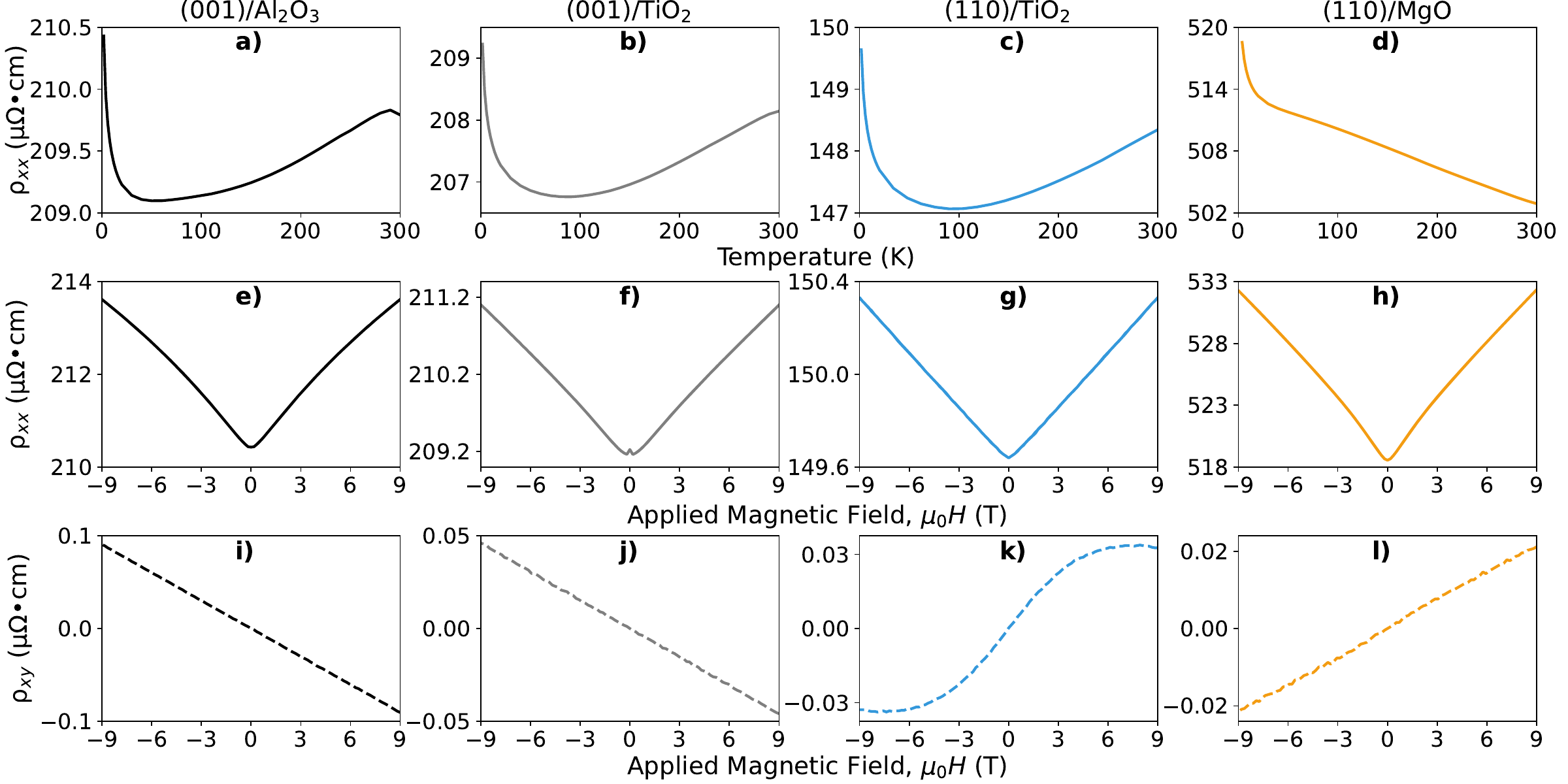}
    
    \caption{\label{Figure 2: PPMS}Electrical characterization of single layer, 1.8~nm thick RuO$_2$ films. (a-d) Temperature dependence of resistivity ($\rho_{xx}$) in zero applied magnetic field. (e-h) Magnetic field dependence of $\rho_{xx}$ at 1.8~K. (i-l) Hall resistivity ($\rho_{xy}$) at 1.8~K. $\rho_{xx}$ and $\rho_{xy}$ are determined through measurements in the van der Pauw orientation on continuous films with an applied current of 100~$\mu$A.}
\end{figure*}

For growth on (001) TiO$_2$/m-Al$_2$O$_3$ and m-Al$_2$O$_3$ substrates, RuO$_2$ films have a preferred (001) orientation, and the (002) reflection is visible in Fig.~\ref{Figure 1: XRD} (b). The expected bulk d-spacing for the (002) reflection is 1.553~\AA, compared to 1.544~\AA~and 1.555~\AA~for the (001)/TiO$_2$ and (001)/Al$_2$O$_3$ samples, respectively, corresponding to out-of-plane strains of $-0.56\%$ and $0.12\%$. The (002) peak of the TiO$_2$ buffer layer is also observed.

Figure~\ref{Figure 1: XRD} (c) shows the magnetization for the 60~nm (001)/Al$_2$O$_3$ sample and a bare sapphire substrate measured at 5~K for an in-plane applied magnetic field. The magnetization is calculated from the measured areas and total thickness of the film plus substrate (no background or substrate subtractions are performed). The observed response is entirely linear, indicating the majority of contribution arises from the diamagnetic Al$_2$O$_3$ substrate, with the RuO$_2$ films showing no detectable magnetic hysteresis or signs of significant magnetization. We report additional background subtracted magnetization in the Supplemental Materials~\cite{SM}.


Figure \ref{Figure 2: PPMS} (a-d) shows resistivity  ($\rho_{xx}$) response from 1.8~nm thick RuO$_2$ films from 300 to 1.8 K. The trend in resistance with decreasing temperature is similar for the (110)/TiO$_2$, (001)/TiO$_2$ and (001)/Al$_2$O$_3$ samples, showing a metallic-like decrease in resistance followed by an increase at lower temperatures. RuO$_2$ films deposited at 1.8~nm are at a metal/insulator boundary, so minor variance between films determine whether films display metallic-like behavior with weak localization at low temperatures or insulating-like behavior with strong localization at low temperatures~\cite{PhysRevMaterials.8.075002}. The (110)/MgO film exhibits a higher resistivity and a temperature trend consistent with being on the other side of the metal/insulator boundary from the other samples. The observed increase in resistivity at low temperatures for all samples is attributed to localization effects~\cite{Jeong2025, PhysRevMaterials.8.075002}. Figure~\ref{Figure 2: PPMS} (e–h) shows the longitudinal resistivity $\rho_{xx}$ as a function of magnetic field at 1.8~K. The magnetoresistance closely matches that reported previously for RuO$_2$ thin films of comparable thickness, exhibiting a positive, smoothly varying field dependence characteristic of conventional orbital transport~\cite{PhysRevMaterials.8.075002}.

Figure \ref{Figure 2: PPMS} (i-l) shows Hall resistivity ($\rho_{xy}$) response from 1.8~nm thick RuO$_2$ films at 1.8~K. For the (001) oriented films, a linear $\rho$$_x$$_y$ is observed with a negative slope, which we attribute to the ordinary Hall component. In contrast, the (110)/TiO$_2$ film shows a distinct field-dependent nonlinear $\rho_{xy}$, which is reversed in sign relative to the (001) oriented samples. Such behavior is consistent with symmetry-allowed anomalous Hall contributions previously reported in (110)-oriented RuO$_2$ films~\cite{Jeong2025}. The (110)/MgO film shows a similar response to the (110)/TiO$_2$ film, albeit with a decrease in magnitude, which we again attribute to an additional anomalous Hall component. Temperature and film thickness dependence of $\rho_{xy}$, and electrical characterization on a second (110)/MgO sample are reported in the Supplemental Materials~\cite{SM}.


\begin{figure}
    \centering
    \includegraphics[width=0.95\linewidth]{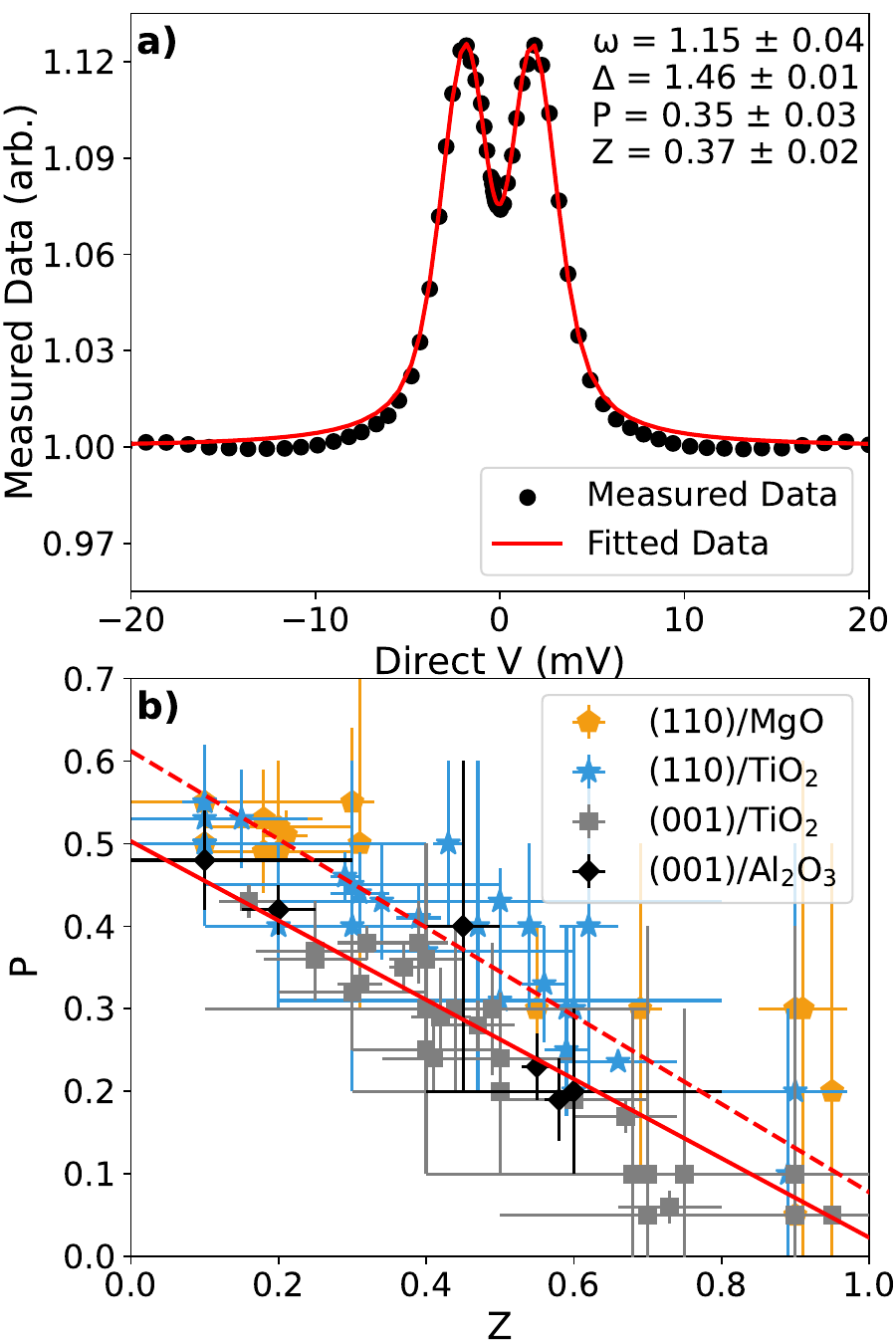}
    \caption{Point contact Andreev reflection measurements on 60~nm RuO$_2$ samples. (a) Exemplar conductance for the (001)/TiO$_2$ sample with best fit to the BKT model (see text). (b) The polarization, P, as a function of barrier strength, Z. The dashed line represents the linear fit of points from (110)-oriented RuO$_2$ films and the solid line represents the linear fit of  points from (001)-oriented RuO$_2$ films.
    \label{Figure 3: PCAR}}
\end{figure}

Spin polarization in the ballistic transport regime can be estimated from PCAR spectroscopy experiments~\cite{soulen1998, baltz2009conductance, PhysRevB.85.064410, Seemann_2010,6971749, PhysRevB.76.174435} by fitting the bias dependence of the conductance with the modified Blonder--Tinkham--Klapwijk (BTK) model~\cite{Strijkers_2001}. Figure~\ref{Figure 3: PCAR} (a) shows a representative processed and fitted PCAR spectrum from the 60~nm (001)/TiO$_2$ sample. The conductance is obtained from bias sweeps between positive and negative voltages and processed by removing a small background contribution and symmetrizing the spectrum prior to fitting with the four-parameter BTK model. Further details of the data acquisition, processing, and fitting procedures are provided in the Supplemental Materials~\cite{SM}.

Figure~\ref{Figure 3: PCAR} (b) shows the dependence of the transport spin polarization, P, as a function of the barrier strength, Z, extracted from 83 spectra taken across all four RuO$_2$/substrate combinations. Separate linear regression lines are shown for the (110)-oriented RuO$_2$ films (dashed line) and the (001)-oriented RuO$_2$ films (solid line). Importantly, the separation in P between the two orientations is evident across the experimentally accessed range of Z, with the (110)-oriented films exhibiting systematically higher P values than the (001)-oriented films at comparable Z. While extrapolation to Z~$= 0$ is commonly used as a phenomenological reference point in PCAR analyses, this limit is not strictly physical. However, the intercepts provide a convenient metric for summarizing the transport spin polarization and its orientation dependence, yielding P~$= 0.61$ for the (110)-oriented films and P~$= 0.50$ for the (001)-oriented films, with a statistical regression uncertainty on the Z~$ \to 0$ intercept of $\pm 0.01$ in each case. These values represent effective transport spin polarizations averaged over the current injection geometry.


A central challenge in establishing altermagnetism in RuO$_2$ has been the apparent discrepancy between bulk-sensitive probes, which generally find no evidence for long-range magnetic order, and thin-film or near-surface measurements that report anomalous transport and spin-split electronic states. Muon spin rotation and neutron diffraction experiments indicate the absence of static magnetic order in bulk RuO$_2$~\cite{kessler2024absence}, whereas anomalous Hall effects, spin-resolved photoemission, and related signatures have been reported primarily in thin films or measurements influenced by reduced symmetry near surfaces or interfaces~\cite{li2025exploration, hussain2025exploring, feng_2022, Jeong2025, akashdeep2025surfacelocalizedmagneticorderruo2}. Within a framework in which altermagnetic electronic states in RuO$_2$ are surface-limited and emerge under such reduced symmetry conditions, this apparent tension can be naturally reconciled: bulk-averaging techniques primarily probe the interior of the material and therefore have limited sensitivity to magnetic states that do not constitute a bulk thermodynamic phase, while transport and local spectroscopic measurements can be influenced by symmetry-broken near-surface regions. 

From a symmetry perspective, such a scenario is compatible with theoretical expectations. Bulk rutile RuO$_2$ is centrosymmetric; however, surfaces and interfaces necessarily break inversion symmetry and reduce the crystalline symmetry that protects spin degeneracy. First-principles studies have shown that symmetry breaking at RuO$_2$ surfaces can stabilize magnetic order and induce spin-polarized electronic states even when the bulk remains nonmagnetic~\cite{6fxv-153y,Brahimi_2025}. In this context, altermagnetism need not manifest as a bulk property, but rather as a surface-limited electronic state.

Our experimental results are consistent with this surface-limited altermagnetism framework in RuO$_2$. The nonlinear anomalous Hall effect is observed only in ultrathin RuO$_2$ films and only for specific crystallographic orientations, while thicker 60~nm films show purely linear Hall responses in $\pm9$~T applied fields (see Supplemental Materials~\cite{SM}), which is consistent with prior reports~\cite{Jeong2025}. This thickness dependence indicates that the electronic states responsible for the anomalous Hall signal are limited to near-surface regions rather than extending throughout the film thickness. Moreover, the strong orientation dependence of the Hall response aligns with symmetry-based predictions for RuO$_2$, where the direction of the N\'eel vector and the resulting transport signatures depend sensitively on crystallographic orientation. Specifically, an anomalous Hall effect is observed only for our (110)-orientation films. This behavior is consistent with theoretical predictions in which the N\'eel vector lies along the easy $\langle001\rangle$ axis in zero field, but rotates toward the $\langle110\rangle$ direction when a magnetic field is applied along the $\langle110\rangle$ direction~\cite{vsmejkal2020crystal,feng_2022}. Such a rotation enables a finite anomalous Hall response for (110)-oriented films, while no anomalous Hall signal is expected for (001)-oriented films under the same conditions.

The PCAR measurements provide complementary evidence for spin-polarized transport in RuO$_2$, despite the absence of detectable net magnetization. Owing to the point-contact geometry and the interfacial nature of Andreev reflection, PCAR is naturally weighted toward electronic states in the vicinity of the contact region~\cite{daghero2010probing}. Andreev reflection at altermagnet/superconductor interfaces is expected to be dominated by quasiparticles near normal incidence, making the measured conductance sensitive to the symmetry and orientation of the spin-split Fermi surfaces~\cite{PhysRevB.108.054511,PhysRevB.108.L060508}.

Experimentally, the extracted transport spin polarization values are systematically higher for (110)-oriented films than for (001)-oriented films. In RuO$_2$, theory predicts that spin-polarized charge transport is maximized along $\langle110\rangle$ crystallographic directions, while transport along $\langle001\rangle$ corresponds to a minimum in the projected transport spin polarization owing to the anisotropic spin splitting of the electronic bands~\cite{PhysRevLett.126.127701}. As a result, in (110)-oriented films the injected current more closely aligns with directions of maximal spin-polarized transport, whereas in (001)-oriented films it predominantly samples directions with reduced spin polarization. Although in both film orientations the in-plane component of the current probes both $\langle001\rangle$ and $\langle110\rangle$ directions, the differing out-of-plane alignment may lead to the higher observed transport spin polarization for (110)-oriented films.

Taken together, the combination of finite transport spin polarization, orientation-dependent anomalous Hall response, and the absence of ferromagnetic hysteresis rules out conventional ferromagnetism as the origin of the observed effects. While non-collinear antiferromagnetic order can, in principle, generate anomalous transport signatures~\cite{Rimmler2025}, such states do not necessarily account for the observed orientation dependence in both Hall and superconducting transport measurements. In contrast, altermagnetism is distinguished by orientation-dependent, symmetry-enforced spin splitting of the electronic structure without net magnetization, providing a natural and unified framework for interpreting both the anomalous Hall effect and the PCAR results~\cite{vsmejkal2020crystal,PhysRevLett.126.127701}.

Several limitations of the present study should be noted. PCAR measurements require conductive films and inherently average over multiple current pathways and crystallographic directions, thereby precluding direct access to the full directional dependence of the spin polarization. However, in the PCAR geometry Andreev reflection is expected to be dominated by quasiparticles near normal incidence, so the extracted $P$ is weighted toward trajectories close to the film normal, unless the transport is strongly anisotropic. Likewise, Hall measurements probe electronic states that may be dominated by near-surface transport in ultrathin films. Neither technique provides a strictly surface- or depth-resolved probe of the electronic structure; rather, both yield transport-weighted information that reflects the regions most relevant to conduction. Future experiments employing directionally selective tunneling spectroscopy, such as the Meservey--Tedrow technique~\cite{PhysRevB.7.318, MESERVEY1994173}, could provide more stringent tests of the altermagnetic electronic structure and its symmetry dependence.

In summary, we have investigated the electronic transport properties of epitaxial RuO$_2$ thin films using anomalous Hall measurements and point-contact Andreev reflection spectroscopy. We observe transport spin polarization and an orientation-dependent anomalous Hall effect in the absence of detectable net magnetization. The (110)-oriented films exhibit higher transport spin polarization across the measured Z range and show anomalous Hall signatures, whereas the (001)-oriented films do not. The restriction of anomalous Hall signatures to ultrathin films indicates that the relevant electronic states are confined near the surface, while the correlated orientation dependence observed in both Hall and superconducting transport measurements is consistent with symmetry-driven spin splitting of the electronic structure. Taken together, these observations are consistent with altermagnetic behavior in RuO$_2$, with the experimentally accessible signatures most likely arising from near-surface regions. More broadly, our results establish superconducting transport spectroscopy as a quantitative probe of spin polarization in compensated magnetic systems and highlight its potential as a metrology for identifying and characterizing altermagnet candidates.

\vspace{12pt}

We acknowledge experimental assistance through the Analysis Research Service Center from Sam Cantrell and Casey Smith, and through the Nanofabrication Research Service Center from Melvin Cruz. S.G. acknowledges support from the UK Engineering and Physical Sciences Research Council through a Doctoral Training Partnership EP/W524372/1 (2828359). G.B. acknowledges support from the UK Engineering and Physical Sciences Research Council through the CAMIE Programme grant EP/X027074/1. N.S. acknowledges support from Texas State University through new faculty startup funding and the Research Enhancement Program. 

\vspace{12pt}

The data that support the findings of this article are openly available~\cite{Data}.

\bibliography{RuO2}

\end{document}


\title{Supplemental Material for ``Transport spin polarization in RuO$_2$ films"}

\author{Alexandra J. Howzen}
\affiliation{Department of Physics, Texas State University, San Marcos, Texas 78666, USA}

\author{Sachin~Gupta}
\affiliation{School of Physics and Astronomy, University of Leeds, Leeds, LS2 9JT, United Kingdom}

\author{Gavin Burnell}
\affiliation{School of Physics and Astronomy, University of Leeds, Leeds, LS2 9JT, United Kingdom}

\author{Nathan~Satchell}
\email{satchell@txstate.edu}
\affiliation{Department of Physics, Texas State University, San Marcos, Texas 78666, USA}

\maketitle

\section{Additional Detail of Substrate and Buffer Layer Preparation}

\textbf{Preparation of the (001) TiO$_2$ buffer layer:} The (001) TiO$_2$ buffer layer on m-plane Al$_2$O$_3$ was prepared by DC sputtering a 60~nm TiO$_2$ film from a 99.999\% pure Ti target at 150~W at a working pressure of 1~mTorr achieved with Ar and O$_2$ flow rates of 10 and 1.1 SCCM, respectively, at 600$^\circ$C. The TiO$_2$ film is then removed from the sputter chamber and annealed in a tube furnace at 800$^\circ$C under an atmospheric pressure of pure O$_2$ flow for one hour.

\textbf{Substrate preparation:} Substrates were cleaned using sonication in IPA then DI water followed by N$_2$ drying prior to loading into the sputter system for thin film growth. Additionally, prior to the growth of the 1.8~nm thick (001)/TiO$_2$ and (110)/TiO$_2$ samples reported in the main text, the substrates (plus buffer layer) were etched in 10\% hydrofluoric acid for 40~s followed by annealing in a tube furnace at 900$^\circ$C for 2 hours under ambient atmosphere.

\section{Further Magnetic Analysis}

\begin{figure}
    \centering
    \includegraphics[width=0.95\linewidth]{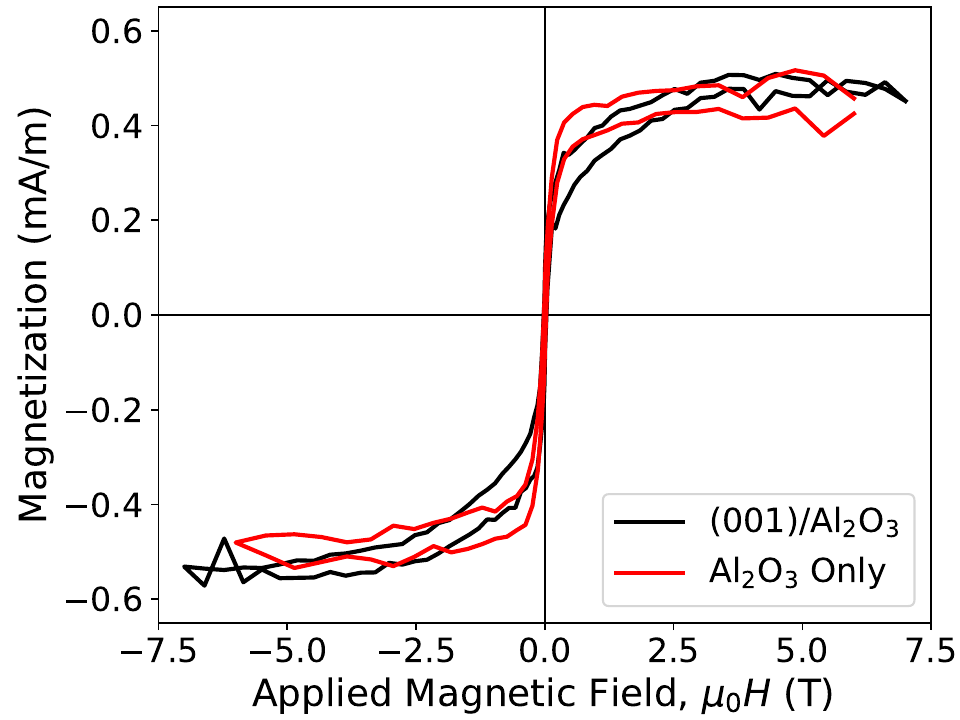}
    \caption{SQUID magnetometry of the film deposited on Al$_2$O$_3$ and a bare wafer comparison with magnetization calculated from the measured areas and the nominal thickness of the film plus wafer. A linear fit was subtracted from the raw data to show any nonlinear response or magnetic remanence. Data acquired at 5~K for a magnetic field applied in-plane on a quartz paddle.}
    \label{fig:S3 Magnetic Remnant SQUID}
\end{figure}

Magnetization data in the main text is presented without the usual substrate subtraction. To further probe for any magnetization in our RuO$_2$ films, we present further analysis here. Figure~\ref{fig:S3 Magnetic Remnant SQUID} shows the magnetization data from the main text following a linear subtraction. A small remanent magnetization is present in measurements of both the (001)/Al$_2$O$_3$ sample and a reference Al$_2$O$_3$ substrate only. The presence of the signal on the substrate only implies that the signal arises from spurious sources within the instrument itself. Such spurious sources of signal are widely reported in prior literature~\cite{10.1063/1.3060808} and do not affect the conclusion that our RuO$_2$ films do not show a magnetic response within the resolution of our magnetometer.

\section{Second Ultrathin $\text{RuO}_2$ Sample on $\text{MgO}$ by DC Sputtering}

\begin{figure}
    \centering
    \includegraphics[width=1\linewidth]{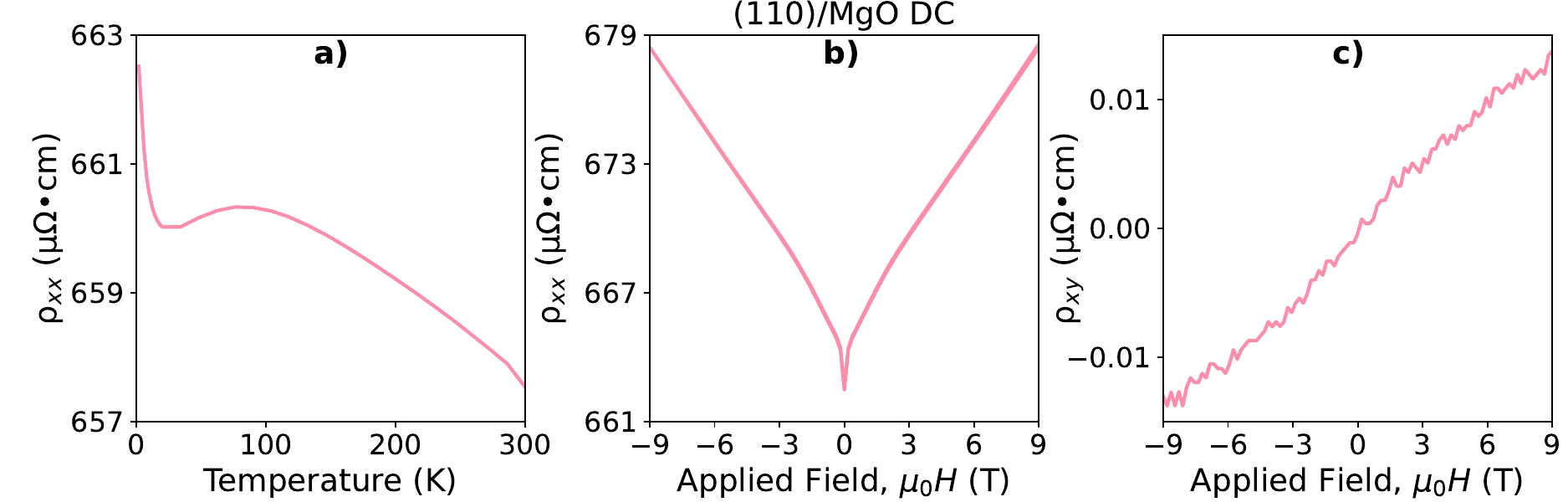}
    \caption{Electrical characterization of single layer, 2.9~nm thick RuO$_2$ film deposited via reactive DC sputtering. (a) Temperature dependence of resistivity ($\rho_{xx}$) in zero applied magnetic field. (b) Magnetic field dependence of $\rho_{xx}$ at 1.8~K. (c) Hall resistivity ($\rho_{xy}$) at 1.8~K. $\rho_{xx}$ and $\rho_{xy}$ are determined through measurements in the van der Pauw orientation on continuous films with an applied current of 100~$\mu$A.}
    \label{fig:S1 MgO DC Film}
\end{figure}

A second (110)/MgO sample was prepared to further investigate the temperature dependence and magnitude of the Hall response for this film/substrate pairing. For this second (110)/MgO sample sample, the MgO substrate was annealed in a tube furnace at 1000$^\circ$C for 40 min in an O$_2$ ambient gas environment following the technique of \cite{10.1063/1.5047029}. Deposition of RuO$_2$ follows the procedure outlined in the main text, where a DC source was used at a power of 32~W, instead of the RF source. Due to a small discrepancy between the RF and DC growth rates in the ultrathin regime, the deposited RuO$_2$ film was approximately 2.9~nm compared to the 1.8~nm thick RF sputtered (110)/MgO sample for the same deposition time (thicknesses were determined via x-ray reflectivity).

Electrical transport characterization of the second 2.9~nm (110)/MgO sample is shown in Figure \ref{fig:S1 MgO DC Film}. The temperature dependence of resistance, shown in Figure~\ref{fig:S1 MgO DC Film} (a), follows a similar increase in resistance with decreasing temperature with a sharp increase in resistivity at low temperatures, as observed in the (110)/MgO sample reported in the main text. Prior to the low temperature increase, there is a decrease in resistivity. Figure~\ref{fig:S1 MgO DC Film} (b) shows the magnetoresistance curve at 1.8~K, where the high field region matches the four main text samples. A distinct V-shape is seen at low magnetic fields and is attributed, following Rajapitamahuni~\textit{et. al}, to the presence of variable range hopping conduction~\cite{PhysRevMaterials.8.075002}. A nonlinear response in the Hall resistivity is observed in Figure~\ref{fig:S1 MgO DC Film} (c), interpreted as the anomalous Hall contribution. In this sample, the magnitude of the anomalous Hall contribution is decreased compared to the (110)/MgO and (110)TiO$_2$ samples reported in the main text. A decrease in anomalous Hall contribution as film thickness increases (2.9~nm compared to 1.8~nm) is consistent with previously reported trends~\cite{Jeong2025}.

\section{Temperature Dependence of Hall Effect}

\begin{figure}
    \centering
    \includegraphics[width=1\linewidth]{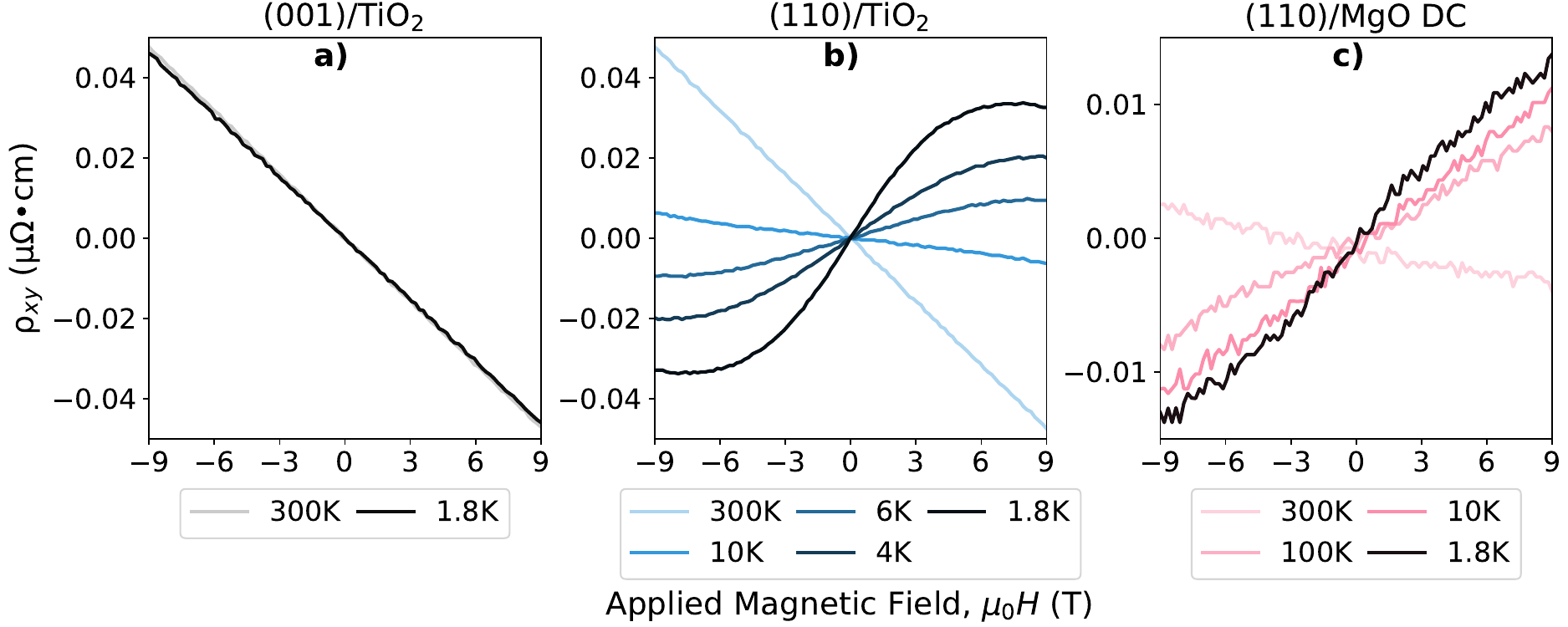}
    \caption{Hall resistivity ($\rho_{xy}$) at 1.8~K. $\rho_{xx}$ and $\rho_{xy}$ are determined through measurements in the van der Pauw orientation with an applied current of 100~$\mu$A for (a) (001)/TiO$_2$, (b) (110)/TiO$_2$, and (c) (110)/MgO DC.}
    \label{fig:S2 T Dependence}
\end{figure}

Figure~\ref{fig:S2 T Dependence} shows the temperature dependence of the Hall resistivity for the (001)/TiO$_2$, (110)/TiO$_2$, and the second 2.9~nm (110)/MgO DC samples. Figure~\ref{fig:S2 T Dependence} (a) demonstrates that there is no temperature dependence of the ordinary hall component observed for the (001)/TiO$_2$ sample. In stark contrast, Figure~\ref{fig:S2 T Dependence} (b,c) show that the anomalous Hall component has a strong temperature dependence for the (110)/TiO$_2$ and (110)/MgO samples. Both of the (110) films show a strong anomalous Hall component at 1.8 K, followed by a decrease in its magnitude with an increase in temperature. At higher temperatures, the sign of the Hall coefficient is reversed completely and by 300~K the Hall resistivity for the (110)/TiO$_2$ sample is linear in field, which is interpreted as the ordinary Hall component only (the anomalous component is no longer present). In particular, comparing the (110)/TiO$_2$ and (001)/TiO$_2$ samples at 300~K, the magnitude of the Hall resistivity is near identical. These samples are both on TiO$_2$ and were grown in the same vacuum cycle. The (110)/MgO DC sputtered film follows the same trend, with a different magnitude due to the different thickness and substrate of this film.

\section{Thickness Dependence of Electrical Properties}

\begin{figure*}
    \centering
    \includegraphics[width=1\linewidth]{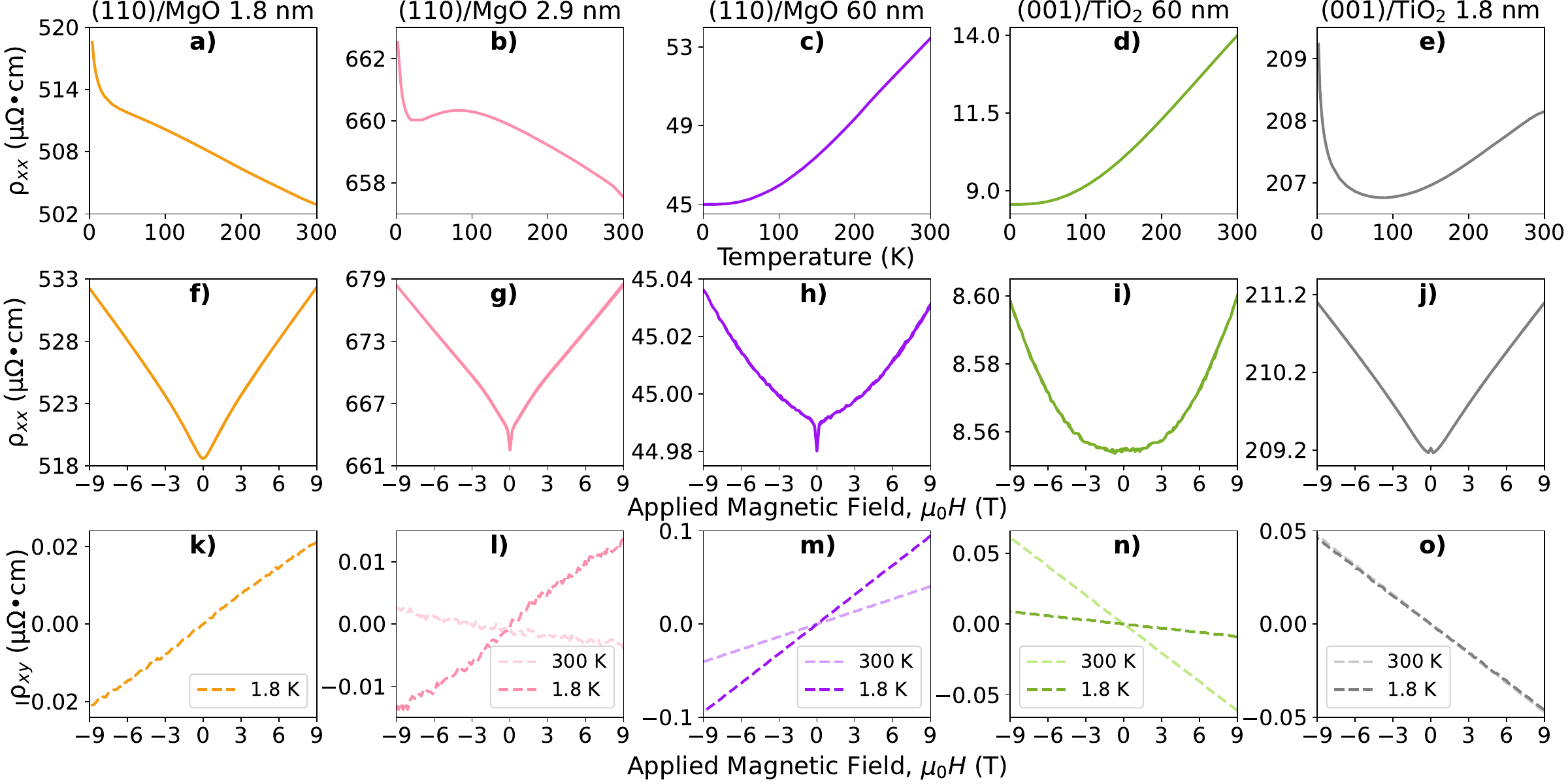}
    \caption{Electrical characterization of 1.8, 2.9, and 60~nm thick RuO$_2$ films. (a-e) Temperature dependence of resistivity ($\rho_{xx}$) in zero applied magnetic field. (f-j) Magnetic field dependence of $\rho_{xx}$ at 1.8~K. (k-o) Hall resistivity ($\rho_{xy}$) at 1.8 and 300~K. $\rho_{xx}$ and $\rho_{xy}$ are determined through measurements in the van der Pauw orientation on continuous films with an applied current of 100~$\mu$A for the 1.8 and 2.9~nm thick films and 10~mA for the 60~nm thick films.}
    \label{Thickness Dependence}
\end{figure*}

Figure~\ref{Thickness Dependence} shows the electrical transport characterization of 1.8, 2.9, and 60~nm (110)/MgO and 1.8 and 60~nm (001)/TiO$_2$ films with some panels reproduced from other figures in this work for convenience. Figure~\ref{Thickness Dependence} (c,d) shows metallic-like temperature dependence of both 60~nm (110)/MgO and (001)/TiO$_2$ films. Figure~\ref{Thickness Dependence} (h,i) shows parabolic magnetoresistance of the same films, with standard $H^2$ dependence at 1.8~K, consistent with expectations for metallic RuO$_2$~\cite{PhysRevMaterials.8.075002}. We note some substrate dependence of the 60~nm films, with the (110)/MgO 60~nm film showing a higher resistivity than the 60~nm (001)/TiO$_2$ film and distinct V-shape magnetoresistance at low magnetic fields, attributed to variable range hopping~\cite{PhysRevMaterials.8.075002}.

Both 60~nm (110)/MgO and (001)/TiO$_2$ films show only a linear Hall response for $\pm9$~T applied magnetic field at both 1.8 and 300~K, Figure~\ref{Thickness Dependence} (m,n). For (110)/MgO, this behavior is distinctly different to the thinner 1.8 and 2.9~nm samples that have nonlinear Hall response at 1.8~K. Based on prior literature, as the film thickness is increased, the nonlinear anomalous Hall response is expected to decrease, with a 9.1~nm film reported by Jeong \textit{et. al} showing a linear Hall response, interpreted to be purely the ordinary Hall contribution~\cite{Jeong2025}. We note our finding that the slope of the ordinary Hall effect is substrate dependent, like the film resistivity and additional low field magnetoresistance features.  

\section{PCAR Data Acquisition and Processing}

The Point Contact Andreev Reflection spectra are measured using an AC lock-in technique, where the differential conductance is determined by subjecting a contact to a controlled DC bias with a small AC modulation that is detected across the contact and a series resistor. The DC bias is typically adjusted over a range of $\pm30$~meV, which is monitored through a separate channel. The bias is swept continuously from zero to positive bias, through zero to negative bias, and back to zero. The PCAR fitting and analysis are performed using the Stoner-PythonCode package~\cite{Burnell2025Stoner}. For analysis, only the portion of the sweep spanning directly between the positive and negative bias extrema and passing through zero bias is retained. A small instrumentation offset applied to the DC bias is also removed during this initial stage of data processing. The data are subsequently decomposed into symmetric and antisymmetric components, and only the symmetric component is retained, as all features of interest in the spectra are expected to be symmetric about zero bias. The BTK model is typically applied to normalized conductance, where normalization is performed using the high-bias conductance regime, in which tunneling or Joule heating effects can produce linear or parabolic backgrounds~\cite{baltz2009conductance}. In this work, the spectra were normalized by dividing by a parabolic fit to the outer 20\% of the bias range.

The modified BTK model used in the analysis of the PCAR data has 4 fitting parameters. Firstly, an interfacial barrier strength $Z$ (dimensionless) which effectively determines a scattering potential at the interface and so will take a small, finite value even for a perfect contact. Secondly, the spin polarization $P=\frac{N_\uparrow v_{f\uparrow}-N_\downarrow v_{f\downarrow}}{N_\uparrow v_{f\uparrow}+N_\downarrow v_{f\downarrow}}$ which thus ranges from 0 for non-spin-polarized materials to 1 for fully spin-polarized materials. Thirdly, the superconducting energy gap $\Delta$ which might not fully match the bulk value of Nb at the tip of the contact due to geometric or physical constraints. Finally, there is a `smearing parameter' $\omega$ that encompasses both thermal effects and athermal scattering processes not otherwise accounted for in the model.

The modified BTK model is then fitted to the now normalized data. We start with an initial set of fitting values of: $Z=0.4$, $\omega=1.5\,$meV, $\Delta=1.25\,$meV and $P=0.4$. However, the initial P and (Z) values were varied from 0.15 (0.85) to 0.55 (0.15) in order to identify and rectify fits with overly large errors in P or Z. A differential-evolution algorithm was utilized to identify a global minimum $\chi^2$ value to locate the best fit. This best-fit location is then used to seed a nonlinear least squares fit using the Levenberg–Marquardt algorithm, allowing for estimated standard errors for the fitting parameters to be determined. Fits with non-physical values of $\Delta$ and excessively large $\omega$ were then discarded as they indicate the presence of an unaccounted spreading resistance~\cite{Woods2004, baltz2009conductance}, before considering all the fitted $(Z,P)$ combinations. However, if reasonable values for $\Delta$ and $\omega$ are attained, but the errors of P and Z were large; individual stepped scans of all fitted parameters for those spectra were performed around the given values to further optimize the fit. Although extrapolation to $Z=0$ is not strictly physical, the $Z\to0$ intercept is commonly used as a phenomenological reference point in PCAR analyses of metallic contacts~\cite{soulen1998, Strijkers_2001}.

A Python script was written to perform the linear regression fitting of the PCAR P vs Z data and extrapolate the spin polarization at the value of Z = 0. The ODR \cite{ODR10.1090} module from the Scipy library  \cite{Virtanen2020} was coupled with a linear model that includes the standard errors when determining the weight of data points in the linear regression to fit the data. The initial conditions were set to -0.5 and 0.55 for the slope and intercept respectively. All Python code utilized in processing, fitting, and plotting the data is available~\cite{Data}.

\bibliography{RuO2}